\documentstyle[prd,aps,eqsecnum,psfig]{revtex}	          

\newcommand{\be}{\begin{equation}}
\newcommand{\ee}{\end{equation}}
\newcommand{\bea}{\begin{eqnarray}}
\newcommand{\eea}{\end{eqnarray}}
\newcommand{\bm}{\bibitem}
\newcommand{\om}{\omega}
\newcommand{\al}{\alpha}
\newcommand{\bt}{\beta}
\newcommand{\lm}{\lambda}
\newcommand{\sg}{\sigma}
\newcommand{\de}{\delta}

\newcommand{\gm}{\gamma}
\newcommand{\ep}{\epsilon}
\newcommand{\Th}{\Theta}
\newcommand{\th}{\theta}
\newcommand{\tq}{\Theta^q}
\newcommand{\tg}{\Theta^g}
\newcommand{\bq}{\bar{q}}
\newcommand{\qq}{\bar{q} q}
\newcommand{\hm}{\hat{m}}
\newcommand{\us}{u \!\!\! /}
\newcommand{\xs}{x \!\!\! /}
\newcommand{\ks}{k \!\!\! /}
\newcommand{\qs}{q \!\!\! /}
\newcommand{\calt}{{\cal T}^{ab}_{\mu\nu}}
\newcommand{\lmn}{\Lambda_{\mu\nu}}    
\newcommand{\vq}{|\vec{q}|}
\newcommand{\la}{\langle}
\newcommand{\ra}{\rangle}

\newcommand{\OP}{\overline{\langle O \rangle}}
\newcommand{\OPp}{\overline{\langle O^{\prime} \rangle}}
\newcommand{\mrs}{m_{\rho}^2}
\newcommand{\frs}{F_{\rho}^2}

\newcommand{\mn}{4m^2_N}
\newcommand{\qz}{q_{0}}

\newcommand{\atm}{\mid_{\mu}}
\newcommand{\atq}{\mid_Q}
\newcommand{\ikx}{ik\cdot x}

\begin{document}

\draft 

\title{
QCD sum rules for $\rho$ mesons in nuclear matter}

\author{S. Mallik\footnote{Electronic address: mallik@tnp.saha.ernet.in}} 
\address{Saha Institute of Nuclear Physics,
1/AF, Bidhannagar, Calcutta 700 064, India} 

\author{Andreas Nyf\/feler\footnote{Electronic address:
nyf\/feler@cpt.univ-mrs.fr}}  
\address{Centre de Physique Th\'{e}orique, CNRS Luminy, Case 907,
F-13288 Marseille Cedex 9, France}

\date{22 May 2001} 

\maketitle

\begin{abstract} 
We investigate QCD sum rules for vector currents in the rho meson
channel in the nuclear medium.  For increased sensitivity, we subtract
out the vacuum contributions.  With a saturation scheme often
considered in the literature, we find these ``subtracted'' sum rules to
be unstable. It indicates the importance of other contributions
neglected in the saturation scheme. These include more Landau
singularities at low energy, additional operators in the medium, and
possibly the in-medium width of the rho meson.
\end{abstract}



\renewcommand{\thefootnote}{\arabic{footnote}}
\setcounter{footnote}{0}


\section{Introduction} 

It was quite some time ago that the original vacuum QCD sum rules
\cite{SVZ} were first extended to a medium at finite temperature and density 
\cite{Shap,Drukarev}. Since then much work has been devoted to 
extracting results from them. Typically one tries to predict the medium 
dependence of the parameters of particles and resonances in  
the low energy region. However, these in-medium sum rules have several 
new features. These must be included quantitatively in the saturation 
scheme for reliable predictions.

One such feature, recognized soon after their initial formulation, is
that the in-medium sum rules have contributions from new operators
over those already present in the vacuum sum rules \cite{Jin,Shuryak}.
The velocity four-vector of the medium allows one to construct
additional Lorentz scalar operators.  (In the rest frame of the medium
they are simply the time components of Lorentz tensor operators.) At
higher dimensions there arises a multitude of such operators.

The other features relate to the hadron spectral function. A
communicating single particle state may acquire a large width (in
addition to a possible shift in mass) due to its scattering from the
particles in the medium, so that it may not be possible to treat it in
the narrow width approximation as in the vacuum. Also multiparticle
states in the medium may give rise to several nearby singularities of
the Landau type.

In this work we reexamine the sum rules in the $\rho$-meson channel
in the nuclear medium at zero temperature. Our saturation scheme is
similar to one mostly employed in the literature, which includes only
some of these features \cite{Hatsuda,Jin_Leinweber,Cohen}. In the
operator product expansion we include properly all the operators up to
dimension 4 and only the four-quark operators at dimension 6. In
the spectral function we take the $\rho$ meson in the narrow width
approximation and the $N\bar{N}$ state.\footnote{Actually there could
be further density independent contributions in the continuum, but
they are eliminated in the subtracted sum rules we shall be
considering.}

A complete or ``full'' in-medium sum rule is actually the corresponding
vacuum sum rule, perturbed by small, density dependent terms. Such a
``full'' sum rule, when evaluated, is guaranteed to be stable, more or
less, irrespective of these small contributions, simply because the
vacuum sum rule is so. To get rid of this insensitivity, we subtract
out the vacuum sum rule from the ``full'' one, retaining on both sides
only the terms to first order in the density.

Though well-known, these sum rules are derived here to bring out some
technical points. We start with a broader framework, namely the
ensemble average of the correlation function of currents at finite
temperature and chemical potential. The special cases of sum rules at
finite temperature and at finite density may then be recovered from
the general formulation by setting respectively the chemical potential
and the temperature to zero. We derive the mixing of two operators of
dimension 4, namely the quark and the gluon parts of the energy
momentum tensor, a result known in the context of deep inelastic
scattering. We also present a field theoretic derivation for the
$N\bar{N}$ contribution to the spectral function.

On evaluation we find that the sum rules as saturated above do not
yield stable results, as we have already reported
recently~\cite{SEWM}. This instability points to the importance of
other contributions we did not include, namely, the Lorentz nonscalar
operators at dimension 6 and the Landau singularities besides the
one from the $N\bar{N}$ state. The neglect of the (large) in-medium width
of the $\rho$-meson also possibly adds to this
instability~\cite{Klingl,Leupold1}.

In Sec.~\ref{sec:kinematics} we introduce the kinematic decomposition
of the two point function of vector currents. In Sec.~\ref{sec:OPE} a
simple derivation of the operator product expansion is presented in
coordinate space.  Then in Sec.~\ref{sec:spec_repr} we construct the
spectral representation from $\rho$ and $N\bar{N}$ intermediate states
at finite density. The sum rules are written and evaluated in
Sec.~\ref{sec:sum_rules}. We conclude in Sec.~\ref{sec:discussion}.


\section{Kinematics}
\label{sec:kinematics} 
 
\noindent
Consider the ensemble average of the time ordered ($T$) product of two
currents
\be \label{two_point} 
\calt (q) = i \int{d^4x} e^{iq\cdot x} \left \langle
T\left\{ V^a_{\mu}(x) V^b_{\nu}(0) \right\} \right \rangle.
\ee
Here $V_{\mu}^a (x)$ is the vector current (in the $\rho$-meson channel)
in QCD,
\be
V^a_{\mu}(x) =\bar{q}(x) \gamma_{\mu}{\tau^a \over 2} q(x), 
\ee
$q(x)$ being the field of the $u$ and $d$ quark doublet and $\tau^a$ the
Pauli matrices. The ensemble average of an operator $O$ is denoted by 
$ \la O \ra$,
\be
\langle O\rangle = {\rm Tr} \left[ e^{-\beta (H-\mu N)} O \right] / 
\ {\rm Tr} \left[ e^{-\beta (H-\mu N)} \right] , 
\ee
where $H$ is the QCD Hamiltonian, $\beta$ is the inverse of the
temperature $T$, and ${\rm Tr}$ denotes the trace over any complete set of
states. Keeping our application in mind, we have introduced the
nucleon chemical potential $\mu$ with the corresponding number
operator $N$.

For explicit calculations one generally chooses the rest frame of the
medium, where the temperature is defined. This breakdown of Lorentz
invariance may be restored if we let the medium have an arbitrary
four-velocity $u_{\mu}$ \cite{Weldon82}.  [In the matter rest frame
$u_{\mu}=(1,0,0,0)$.]  The time and space components of $q_{\mu}$ are
then raised to the Lorentz scalars, $w = u\cdot q$ and
$\bar{q}=\sqrt{w^2-q^2}$.

There are two independent, conserved kinematic covariants
\cite{Kapusta}, in which the two point function from
Eq.~(\ref{two_point}) may be decomposed. In choosing their forms we must
note that the dynamical singularities extend up to $q^2=0$. We thus
have to ensure that the kinematic covariants do not have any
singularity in this region, particularly at $q^2=0$. The covariants
$P_{\mu\nu}$ and $Q_{\mu\nu}$ defined by
\be
P_{\mu\nu}=-g_{\mu\nu} +{q_{\mu}q_{\nu}\over q^2} -{q^2\over \bar{q}^2}
\tilde{u}_{\mu} \tilde{u}_{\nu}, \qquad Q_{\mu\nu}={q^4\over \bar{q}^2}
\tilde{u}_{\mu}\tilde{u}_{\nu},
\ee
where $\tilde{u}_{\mu} = u_{\mu} - w q_{\mu}/ q^2$, are indeed so, 
as can be seen by inspecting their components for space and time indices.
The invariant decomposition of ${\cal T}^{ab}_{\mu\nu}$ is then given
by
\be
{\calt} (q) =\de^{ab} (Q_{\mu\nu} T_l + P_{\mu\nu} T_t),
\ee
where the invariant amplitudes $T_{l,t}$ are functions of $q^2$ and $w$.

The kinematic covariants are still not regular: they have a dependence
on the direction of the three-vector $\vec{q}$ even as $\vq
\rightarrow 0$. Thus the spatial components of $\calt$ in this limit
become
\be
{\cal T}_{ij}^{ab}(q) 
= \de^{ab}[\de_{ij} T_t - n_i n_j (T_t -q^2_0 T_l)],
\ee
where $n_i$ is the $i$th component of the unit vector along $\vec{q}$.
To eliminate this direction dependence, we must require that
\cite{Shap}
\be
T_t(q_0, \vq =0)=q_0^2 T_l(q_0,\vq=0).
\ee


\section{Operator product expansion}
\label{sec:OPE} 

Here we outline the derivation of the short distance expansion of the
product of two current operators in Eq.~(\ref{two_point}). As already
stated, the availability of the velocity four-vector $u^\mu$ of the
medium allows one to construct new, independent Lorentz scalar
operators.  Thus up to dimension 4 we have, in addition to the old
ones $ \ {\mathbf 1},\ \qq =\bar{u}u +\bar{d}d, \ G^2 \equiv (\al_s/\pi)
G_{\mu\nu}^a G^{\mu\nu a},$ appearing in the vacuum expectation value,
two new ones $\tq \equiv u^{\mu}\tq_{\mu\nu}u^{\nu}$ and $\tg \equiv
u^{\mu}\tg _{\mu\nu}u^{\nu}$ for the ensemble average. Here
$G_{\mu\nu}^c, \ c=1,...,8$ are the gluon field strengths and $\al_s
=g_s^2/4\pi$, $g_s$ being the QCD coupling constant.
$\Th^{q,g}_{\mu\nu}$ are the quark and the gluon parts of the
traceless energy momentum tensor
\bea
\Th_{\mu\nu}^q &=& \bar{q}i\gm_{\mu}D_{\nu}q -{\hm\over
4}g_{\mu\nu}\qq, \nonumber \\
\Th_{\mu\nu}^g &=&  -G_{\mu\lm}^c G^{~\lm c}_{\nu} +{1\over 4}g_{\mu\nu}
G_{\al\bt}^c G^{\al\bt c}, 
\eea
where $\hm$ is the quark mass in the limit of SU(2) symmetry.

We now obtain the (singular) coefficients of the operators in
coordinate space \cite{Fritzsch,Hub,Mallik97,Novikov}.  In the free
field theory these are obtained from the Wick decomposition of the
operator product. If we treat the gauge fields as external, the same
form of decomposition also holds for the interacting theory
\bea
T \left\{ V^a_{\mu}(x)V^a_{\nu}(0) \right\}& = &
{\rm tr}[\gm_{\mu}S(x,0)\gm_{\nu}S(0,x) \tau^a\tau^b/4] \nonumber \\
& &-i\bar{q}(0)\gm_{\nu}S(0,x)(\tau^b\tau^a/4)q(x) 
-i\bar{q}(x)\gm_{\mu}S(x,0)\gm_{\nu}(\tau^a\tau^b/4)q(0) \nonumber \\
& &+ {\rm a \ regular \ term}.
\label{Wick} 
\eea
The trace in the first term is over all the indices, namely spin,
flavor and color of the quark field. The quark propagator $S(x,0)$ is
now in the presence of the (matrix valued) gauge potential $A_{\mu}(x)$,
\be \label{quark_propagator}
\left\{ -i\gm^{\mu}[\partial_{\mu} -igA_{\mu}(x)]+\hm \right\} S(x,0) =
\de^4(x). 
\ee

An important technical point here is that quantities can be put in a gauge
covariant form from the beginning, if we use the Fock-Schwinger gauge, 
in which, $x^{\mu}A_{\mu}(x) = 0$. In this gauge
\bea
A_{\mu}(x) & = & (1/2)x^{\al}G_{\al\mu}(0) + \cdots \, , \nonumber \\
q(x) & = & q(0) + x^{\mu}D_{\mu}q(0) + \cdots  \, . 
\eea
Equation~(\ref{quark_propagator}) may be solved for the quark
propagator in a series of increasing number of gluon field strengths.

The unit and the quark operators $\; {\mathbf 1}, \; \qq , \;$ and $\;
\tq \;$ reside in the first and the next two terms in
Eq.~(\ref{Wick}), respectively. To calculate their coefficients, it
suffices to use the free quark propagator
\be
S_0(x,0) = {1\over {4\pi^2}} \left(- {2\xs\over {(x^2-i\ep)^2}}
           -{i\hm\over {x^2-i\ep}} \right),
\ee
where we retain only the leading singular term in $\hm$.  The quark
operators in Eq.~(\ref{Wick}) can then be projected on to the scalar
operators by using the formulae
\bea
\la \bq ^j_Aq^k_B\ra & = & {1\over 8} \de^{jk} (\de_{BA} \la \bq q\ra +
(\us)_{BA} \la\bq\us q\ra ), \\ 
\la \bq ^j_Ai(D_{\mu}q)^k_B\ra & = & {1\over 2}\de^{kj} \left\{
{1\over 4}\de_{BA} u_{\mu} \hm \la \bq \us q\ra + (\gm_{\mu})_{BA}
\left ({\hm\over 16}\la \bq q\ra -{1\over 12}\la \tq \ra \right )
+u_{\mu}(\us)_{BA} {1\over 3} \la\tq\ra 
\right\} . 
\eea
The flavor and spinor indices of the quark field have been denoted by
$(j,k)$ and $(A,B)$, respectively. A sum over color indices on both
sides is understood. These projection formulae are easily obtained by
expanding the quark bilinear matrix in terms of the complete set of
Dirac matrices, noting parity conservation, and using the free equation
of motion for the quark field. (Chirality forbids $\la \bq \us q\ra$
to appear in the vector-vector correlation function.)

The two-gluon operators $G^2$ and $\tg$ arise only from the first term
in Eq.~(\ref{Wick}). To get their coefficients we have to work out the
quark propagator up to two gluons.  The general fourth rank tensor
with two gluon fields may be projected on to the scalar operators by
\bea
\la {\rm tr}^c G_{\al\bt}G_{\lm\sg}\ra &=& {1\over
6}(g_{\al\lm}g_{\bt\sg}-g_{\al\sg} 
g_{\bt\lm})\left ( {1\over 4}\la G_{\mu\nu}^c G^{\mu\nu c}\ra +\la\tg\ra
\right ) \nonumber \\
 & &-{1\over 3}(u_{\al}u_{\lm}g_{\bt\sg}-u_{\al}u_{\sg}g_{\bt\lm}
-u_{\bt}u_{\lm}g_{\al\sg}+u_{\bt}u_{\sg}g_{\al\lm}) \la\tg\ra, 
\label{projection} 
\eea
${\rm tr}^c$ indicating trace over the color matrices. It turns out that 
if we insert this expression into the two gluon terms present in $S$,
they do not contribute to $G^2$. Thus the calculation of its coefficient
reduces to inserting $S$ up to the one-gluon term
\be
S(x,0)=S_0(x,0) 
-{ig_s \gm^{\mu} \xs \gm^{\nu}G_{\mu\nu}\over 16\pi^2(x^2-i\ep)},
\ee
in the first term of Eq.~(\ref{Wick}) and then using
Eq.~(\ref{projection}).  However the two-gluon terms in $S$ do
contribute to $\tg$. But it is not necessary to calculate this
coefficient. As we show below, the renormalization group improvement
automatically brings it in.

Having indicated the procedure to calculate the coefficients in coordinate
space, we take the Fourier transform and write the result in momentum
space $(Q^2 \equiv - q^2)$, 
\bea
\lefteqn{\hspace*{-1.2cm} i \int{d^4x} e^{iq\cdot x}T \left\{
V^a_{\mu}(x)V^b_{\nu}(0) \right\} =\de^{ab} \left[ Q_{\mu\nu} \left\{
-{1\over {8\pi^2}} \ln(Q^2/{\mu}^2) {\bf 1}
+{1\over Q^4} \left( {1\over 2}\hm\qq +{1\over 24}G^2 + {2\over 3}\tq\right)
\right\}\right. } \nonumber \\
&& \left. + P_{\mu\nu} \left\{ {Q^2\over {8\pi^2}} \ln(Q^2/{\mu}^2) {\bf 1}
-{1\over Q^2} \left( {1\over 2}\hm\qq +{1\over 24}G^2 + 
{2\over 3} \left(1-{2\bar{q}^2\over Q^2 }\right)\tq\right)\right\}
\right] . 
\label{Fourier_trafo} 
\eea
Here $\mu$ is the renormalization scale, which we take to be $1~{\rm
GeV}$ as the natural scale to normalize the operators. But this scale
is not convenient for the coefficients.  These, calculated above only
to the lowest order in $\al_s$, may contain powers of $\ln(Q^2/\mu^2)$
in higher orders. To get rid of these large logarithms,~\footnote{The
logarithm in the coefficient of the unit operator in
Eq.~(\ref{Fourier_trafo}) is of different origin. See the first
footnote in Sec.\ 4.6 of Ref.~\cite{SVZ}.} we must shift the
renormalization point for the coefficients from $\mu^2$ to $Q^2$.

Now the left hand side of Eq.~(\ref{Fourier_trafo}) is independent of
the renormalization scale.  So are all the operators ${\mathbf 1},
\hm\qq, G^2$ and hence their coefficients. But it is not so for $\tq$
and $\tg$.\footnote{If we had calculated the coefficient of $\tg$, we
would have found it to be proportional to $\al_s
\ln(Q^2/\mu^2)$.  It signals mixing with other operators.} Indeed, it
is well-known in the context of deep inelastic scattering that $\tq$
and $\tg$ mix under a change of scale \cite{Peskin}.  Let us denote
the contribution of these operators generally as
\be
C_q\atm  \tq\atm + C_g\atm \tg\atm \equiv \sum_i C_i\atm \Th^i\atm ,
\ee
where only the dependence on the scale $\mu$ is shown explicitly and
the index $i$ runs over $q$ and $g$. The coefficients $C_i$
satisfy a coupled set of renormalization group equations having the 
solution
\be \label{sol_RG_eq} 
C_i\atm = C_j\atq \left ( e^{-\int_0^t \gm (t')dt'} \right )_{ji},
\ee
where $t={1\over 2} \ln(Q^2/{\mu}^2)$ and $\gm$ is the $2\otimes 2$
anomalous dimension matrix. In the basis we are using it is
\be 
\gm = - {g_s^2\over {(4\pi)^2}} 
\left( \begin{array}{cc} 
-{64\over 9}  &  {4\over 3} n_f \\
 {64\over 9}  &  -{4\over 3}n_f 
\end{array} \right) , 
\ee
where $n_f (=2)$ is the number of flavors. The exponentiation in
Eq.~(\ref{sol_RG_eq}) can be easily worked out by a spectral
decomposition of the matrix $\gm = \sum_i \lm_i P_i$, where $\lm_i$
and $P_i$ are the eigenvalues and the corresponding projection
operators \cite{Gross}. The eigenvalues are 
\be
\lm_1 = 0, \quad \lm_2= {g_s^2 \over (4\pi)^2} {4 \over 3} \left( {16
\over 3} + n_f \right) , 
\ee
and the projection operators can be built out of the (right)
eigenvectors and left eigenvectors. We thus finally get
\be
\sum_i C_i\atm \Th_i\atm = {1\over \left({16\over 3} + n_f\right)}
\left\{ \left. \left(n_f C_q + {16\over 3} C_g\right) \right\vert_Q
\Th +\lm (Q^2) (C_q -C_g)\atq \left. \left({16\over 3} \tq -n_f\tg
\right) \right\vert_Q \right \} ,  
\ee
where $\Th=\tq+\tg$ and the anomalous dimension of the second operator
resides in $\lm (Q^2)$, 
\be
\lm (Q^2)=\left ( {\al_s(\mu^2)\over \al_s(Q^2)} \right ) ^{-d/2b},
\qquad
d={4\over 3} \left({16\over 3} + n_f \right), \qquad b =11- {2\over
3} n_f. 
\ee
We will use $\alpha_s(\mu^2 = 1~{\rm GeV}^2) = 0.4$ in our numerical
evaluations below.

Once the renormalization group improvement is made, we can replace the
coefficient functions by their lowest order values. These may be read
off from Eq.~(\ref{Fourier_trafo}) for the amplitude $T_l$, for
example, as $C_f\atq =2/(3Q^4), \ C_g\atq =0$. Thus the
renormalization group improved result for $T_l$ reads (with $n_f=2$)
\be
T_l = -{1\over 8\pi^2}\ln(Q^2 / \mu^2)
 + {1\over Q^4}\left( {1\over 2}\hat{m}\la \qq \ra + {\la G^2\ra\over 24}
 + {2\over 11}\left\{ \la \Th \ra +\lm (Q^2)\left\la {8 \over 3}
\Th^q-\Th^g\right\ra\right\} \right). 
\ee
Notice that if $\lm(Q^2)$ is set equal to unity, it collapses to the
previous result in Eq.~(\ref{Fourier_trafo}). The amplitude $T_t$ is
also given by the same expression, except for an overall factor of
$-Q^2$ and a factor of $(1-2\bar{q}^2/ Q^2)$ multiplying the term with
$\Th$'s.

Let us summarize the effect of mixing of $\tq$ and $\tg$ under the
renormalization group. It introduces the multiplicative factor, $\lm
(Q^2)$, which is nonperturbative in the sense that it is a sum over
all the leading logarithms in the perturbation expansion. It is true
that the coefficient of $\tg$ is smaller than $\tq$ by a factor of
$g_s^2$. We have left out this smaller coefficient, as did the earlier
authors, but retained the nonperturbative factor. Numerically the
factor is significant for $Q^2$ away from $\mu^2 = 1~{\rm GeV}^2$.

At dimension 6 we include only the Lorentz scalar operators in vacuum,
namely the four-quark operators \cite{SVZ}. In the approximation of ground
state saturation, its contribution  reduces to the familiar expression
involving $\la \bar{q} q\ra^2$.


\section{Spectral representation}
\label{sec:spec_repr}

QCD sum rules make contact with the physical states through the
spectral representation of the two point function. For the vacuum
amplitudes, these are the K\"{a}llen-Lehmann representations in $q^2$,
the only variable available there. At finite temperature and chemical
potential, it is convenient to use the Landau representation, which is
a spectral decomposition in $q_0$ at fixed $\vq$ \cite{Landau}.  For
the invariant amplitudes they are
\be
T_{l,t}(q_0, \vq) = {1\over{\pi}} \int^{+\infty}_{-\infty}dq^{\prime}_0 
{{\rm Im} T_{l,t}(q^{\prime}_0, \vq)\over{q^{\prime}_0 - q_0 - i\ep}}
{\rm tanh}(\bt \qz^\prime/2), 
\ee
up to subtraction terms. The imaginary parts from different
intermediate states can be written as phase space integrals over the
matrix elements of currents. But we shall not write these expressions
directly.  Instead, we first calculate the full amplitudes with these
intermediate states in the real time formulation of field theory at
finite temperature and density \cite{Niemi} and then read off their
imaginary parts. Since we have to calculate the amplitudes at the tree
level, only the $11$-component of the propagators will be needed.

\begin{figure}[!t]
\centerline{\psfig{figure=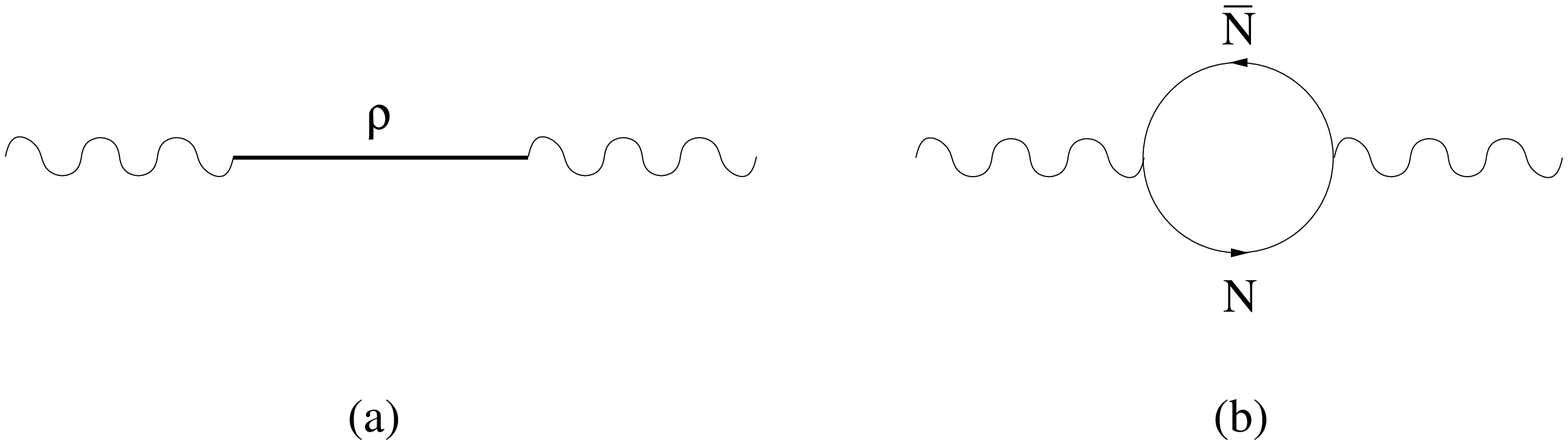,width=14cm,height=4cm}}
\caption{Contributions to the spectral function from (a) the $\rho$
meson and (b) the $N \bar{N}$ state.} \label{fig:Feynman_diagrams}
\end{figure}

We first consider the $\rho$-meson pole, see
Fig.~\ref{fig:Feynman_diagrams}(a), which would dominate the 
absorptive part of the vacuum correlation function. 
\noindent
In the vacuum, its coupling to the current $V^a_{\mu}(x)$ is given by
\be
\langle0|V_{\mu}^a (0)|\rho^b\rangle = \de^{ab} F_{\rho} m_{\rho}
\epsilon_{\mu}, 
\ee
where $\ep_{\mu}$ and $m_{\rho}$ are the polarization vector and the
mass of the $\rho$. Experimentally $F_{\rho} = 153.5~{\rm MeV}$. This matrix
element suggests that, as far as the $\rho$-meson contribution is
concerned, we may write the operator relation
\be \label{operator_relation} 
V^a_{\mu}(x) = m_{\rho}F_{\rho}\rho^a_{\mu}(x).
\ee
In the medium such a relation is generally modified in two
ways. First, the parameters $m_{\rho}$ and $F_{\rho}$ must be replaced
by their effective values in the medium, to be denoted below by a
star. Secondly, the space and the time components have different
constants of proportionality. However, the latter modification takes
place in higher orders in density and temperature
\cite{Leutwyler94}. As we work to the lowest order, the operator
relation~(\ref{operator_relation}) changes in the medium simply to
\be
V^a_{\mu}(x) = m_{\rho}^{\star}F_{\rho}^{\star}\rho^a_{\mu}(x).
\ee
Then the $\rho$-meson contribution to the correlation function is
given essentially by its propagator in the medium. As before, we get
\be
\left( \begin{array}{cc} {\rm Im} T_l(q_0, \vq^2) \\ {\rm Im} T_t(q_0,
\vq^2) \end{array}\right) =   
\left( \begin{array}{cc} 1 \\ m_\rho^{\star 2}\end{array}\right) 
\pi F_{\rho}^{\star 2} \de \left( q_0^2 - (\vq^2 + m_\rho^{\star 2})
\right) {\rm coth}(\bt q_0/2) . 
\ee

In the medium there are other equally important contributions. These
are from the $2\pi$ and, if the medium is nuclear, also the $N\bar{N}$
intermediate state, see Fig.~\ref{fig:Feynman_diagrams}(b). As we
restrict in this work to sum rules at finite nuclear density but at
zero temperature, the $2\pi$ contribution would be small and close to
its vacuum value.  The $N\bar{N}$ contribution in vacuum would also be
small, beginning with a cut at the threshold $q_0^2=4 m_N^2
+\vq^2$.  However, in the nuclear medium the current also interacts
with the real nucleons to give rise to a short cut around the origin,
$-\vq< q_0 <+\vq$. Below we evaluate this contribution.

The contribution of the nucleon isodoublet, spinor field $N^i_A (x), \
i=1,2,\ A=1,...,4,$ to the vector current is
\be
V_{\mu}^a(x) = \bar{N} (x) \gm_{\mu}{\tau^a\over 2} N(x).
\ee
The $N\bar{N}$ contribution to the two point function is then
\be \label{NbarN_contribution} 
\calt (q) = {i\over 2} \de^{ab} \int d^4x e^{iq\cdot x} {\rm tr}
[\gm_{\mu} {\cal S}(x)\gm_{\nu} {\cal S}(-x)] ,
\ee 
where the trace is over the gamma matrices and $i{\cal S}(x)$ is the
11-component of the nucleon propagator in the medium $\la T \left\{ 
N^i_A(x)\bar{N}^j_B(y) \right\} \ra = i\de^{ij} {\cal S}_{AB}(x-y)$, with
\be
i{\cal S}(x-y) = \int {d^4k\over {(2\pi)^4}} e^{-ik\cdot (x-y)} (\ks +m_N)
\Bigg\{ {i\over {k^2-m_N^2+i\ep}} -2\pi\de(k^2-m_N^2)\{n^{-} \th (k_0) 
+ n^{+}\th(-k_0)\}\Bigg\}, 
\ee
the distribution functions $n^{\mp}$ for the nucleon and the
anti-nucleon being given by $n^{\mp}(\om) = (e^{\bt (\om \mp \mu)}
+1)^{-1}$ with $\om =\sqrt{\vec{k}^2 +m_N^2}$.

Our aim is to integrate out the time component of the loop momentum, when
the imaginary part of $\calt (q)$ can be read off from the resulting
expression. However, the above expression for the propagator is not
convenient for this purpose because of the presence of $\th(\pm k_0)$.
So we carry out the $k_0$ integration in the propagator itself,
getting
\bea
i{\cal S}(x) = \int {d^3k\over {(2\pi)^3 2k_0}} & & \left [ 
\th(x^0) \left \{ (\ks
+m_N)(1-n^-)e^{-\ikx} -(-\ks +m_N)n^+ e^{\ikx}\right\} \right .\nonumber \\
& + & \left . \th(-x^0) \left \{-(\ks+m_N) n^- e^{-\ikx} +(-\ks
+m_N)(1-n^+)e^{\ikx}\right\} \right ] . 
\eea
Here the time component of $ k_{\mu}$ is understood to be given by
$\om$.  We can now integrate over $x^{\mu}$ in
Eq.~(\ref{NbarN_contribution}) to get $\de$-functions in three-momentum
and energy denominators in the time components. After some algebra we
get for the imaginary part
\be
{\rm Im}\calt (q) = -{\pi\over 2}\de^{ab} {\rm coth}(\bt q_0/2)
\{I_{\mu\nu}(q) +  I_{\mu\nu}(-q)\} , 
\ee
where
\be \label{I_munu}
I_{\mu\nu}(q) =  \int{d^3k\over (2\pi)^3} {\lmn\over
{4\om_1\om_2}} [(1 - n_1^- - n_2^+)\de(q_0 - \om_1 - \om_2) - (n_2^- -
n_1^-) \de(q_0 - \om_1 + \om_2)],
\ee
with
\bea
\lmn &\equiv & {\rm tr}\{\gm_{\mu}(\ks + m_N)\gm_{\nu}(\ks-\qs+m_N)\} ,
\nonumber \\
     &=& 4[2k_{\mu}k_{\nu}-(k_{\mu}q_{\nu}+k_{\nu}q_{\mu})
-g_{\mu\nu}\{k\cdot(k-q)-m_N^2\}] . 
\eea
Here the subscripts on $n^{\mp}$ serve to indicate that their
arguments are $\om_1 =\sqrt{\vec{k}^2 +m_N^2}$ and $ \om_2 =
\sqrt{(\vec{k}-\vec{q})^2 +m_N^2}$, respectively.  The first and the
second terms in Eq.~(\ref{I_munu}) are nonvanishing in the timelike
($q^2>4m^2_N$) and spacelike ($q^2<0$) regions, respectively.  Working
out the angular integration, it becomes
\bea
I_{\mu\nu}(q)&=&{1\over 4|{\vec q}|}\int^{\om_+}_{\om_-}{d\om_1\over
(2\pi)^2} \lmn (1 - n_1^- - n_2^+)\theta(q^2 - \mn)\nonumber\\
& &- {1\over 4|{\vec q}|}\int^{\infty}_{\om_+}{d\om_1\over (2\pi)^2}
\lmn (n_2^- - n_1^-)\theta(-q^2),
\eea
where the integration limits are given by $\om_{\pm}=[q_0\pm |\vec{q}|
v(q^2)]/2, \ $ with $v(q^2)=\sqrt {1-{\mn/ q^2}}$.

To project the tensor $I_{\mu\nu}$ on the imaginary parts of the
invariant amplitudes it is convenient to form the scalars
$I^{\mu}_{\mu}$ and $u^{\mu} I_{\mu\nu} u^{\nu}$ and relate them to
the invariant amplitudes. Changing the variable $\om_1$ to $x$ given
by $\om_1={1\over 2}(q_0+\vq x)$, we finally get for the contribution
from the $N{\bar N}$ state  
\be
\left(\begin{array}{c} {\rm Im} T_l(q_0, \vq^2) \\ {\rm Im}
T_t(q_0,\vq^2)  \end{array}\right) =        
- {\pi \over 2} \coth(\beta q_0 /2) 
\left[ - {v(3-v^2)\over 24\pi^2}
\left(\begin{array}{c}1\\q^2\end{array}\right)  
+ \left(\begin{array}{c} I_l^+ \\ I_t^+ 
 \end{array}\right) \right] , 
\qquad {\rm for}\ \  q^2>\mn , 
\ee
with
\be \label{I_plus} 
\left(\begin{array}{c} I_l^+ \\ I_t^+ 
 \end{array}\right) = {-1\over 32\pi^2}\int^{v}_{-v} dx
\left(\begin{array}{c}{2(x^2-1)}\\-q^2(2-v^2+x^2)
\end{array}\right)\{n((q_0 +\vq x)/2-\mu)-n((q_0 - \vq x)/2 +\mu)\} , 
\ee
and 
\be
\left(\begin{array}{c} {\rm Im} T_l(q_0, \vq^2) \\ {\rm Im}
T_t(q_0,\vq^2) \end{array}\right) = - {\pi \over 2} \coth(\beta q_0 /2)
\left(\begin{array}{c} I_l^- \\ I_t^- 
\end{array}\right) , 
\qquad {\rm for}\ \  q^2 \leq 0 , 
\ee
with 
\be \label{I_minus}
\left(\begin{array}{c} I_l^- \\ I_t^- 
 \end{array}\right) = 
{- 1\over 32\pi^2}\int^{\infty}_v dx 
\left(\begin{array}{c}{2(x^2-1)}\\-q^2(2-v^2+x^2) 
\end{array}\right) \{n((-q_0 +\vq x)/2 -\mu)-n((q_0 +\vq x)/2-\mu)\}
. 
\ee 
The superscripts $(\pm)$ on $I_{\l,t}$ denote timelike and spacelike
$q_\mu$, respectively.  The distribution function $n$ appearing in
Eqs.~(\ref{I_plus}) and (\ref{I_minus}) is defined by $n(\varepsilon)
= (e^{\beta \varepsilon} + 1)^{-1}$. 

Still higher continuum contributions, at least their leading parts,
are density independent, as is shown by the coefficient of the unit
operator. They will drop out in our subtraction process.


\section{Sum Rules}
\label{sec:sum_rules}

We now go to the spacelike region in $q_{\mu}, (Q_0^2=-q_0^2 >0 )$ at
fixed $\vq$ and take the Borel transform of both the spectral
representation and the operator product expansion with respect to
$Q_0^2$. The sum rules are obtained by equating these transforms at
sufficiently high $M^2$. For $T_l$ it is
\bea
& & F^{\star 2}_\rho e^{-m_\rho^{\star 2}/M^2} + 
{1\over {48 \pi^2}} \int_{\mn}^\infty dq_0^2 e^{-{q_0^2/M^2}}
v^3(q_0^2 +\vq^2)\nonumber\\
&+& e^{\vq^2/M^2}\left( \int_{\mn+{\vq}^2}^\infty dq_0^2 
e^{-{q_0^2/M^2}} I_l^+(q_0,\vq) +
\int_0^{{\vq}^2}dq_0^2 e^{-{q_0^2/M^2}} 
I_l^-(q_0,\vq)\right)\nonumber\\
&=&  {M^2\over 8\pi^2} + {\la O \ra\over M^2} 
- {\la O^{\prime}\ra\over{2M^4}} , \label{SR_1}  
\eea
where $\la O \ra$ and $\la O^{\prime}\ra$ represent the contributions of all
dimension 4 operators and of dimension 6, scalar operators 
in vacuum, respectively,
\be
\la O \ra = {1\over 2} \hat{m} \la \qq \ra + {\la G^2 \ra \over 24}+
{2\over 11}\left\{\la \Th \ra + \lm(M^2)  \left\la {8 \over 3} \Th^q
-\Th^g \right\ra \right\}, \quad \la O^{\prime}\ra = {7g_s^2\over 81}
{\la \bar{q}q\ra}^2. 
\ee
The amplitude $T_t$ satisfies a similar sum rule.

Considerable simplification results if we take $\vq \rightarrow 0$.
This limit may be taken inside the integrals in Eq.~(\ref{SR_1}),
except for the last one, where the integrand diverges, while the range
of integration shrinks to zero. A careful evaluation gives a result
resembling a pole at $q^2=0$.\footnote{The situation is similar to
that at finite temperature \cite{Shap}. See also
Ref.~\cite{Mallik_Mukherjee_1}.} Then the above sum rule for $T_l$
becomes
\bea
& & F^{\star 2}_\rho e^{-m_\rho^{\star 2}/M^2} + {1\over {48 \pi^2}}
\int_{\mn}^\infty dq_0^2 v_0(3-v_0^2)
\left[ n(q_0/2-\mu) + e^{-{q_0^2/M^2}}\{1-n(q_0/2-\mu)-n(q_0/2+\mu)\} 
\right] \nonumber\\ 
&=&  {M^2\over 8\pi^2} + {\la O \ra\over M^2} - {\la
O^{\prime}\ra\over{2M^4}},  
\eea
while the sum rule for $T_t$ in this limit may similarly be found to
be
\bea
& & m_{\rho}^{\star 2}F^{\star 2}_\rho e^{-m_\rho^{\star 2}/M^2} + 
{1\over {48 \pi^2}} \int_{\mn}^\infty dq_0^2 q_0^2 v_0(3-v_0^2) 
e^{-{q_0^2/M^2}}\{1-n(q_0/2-\mu)-n(q_0/2+\mu)\}
\nonumber\\
&=&  {M^4\over 8\pi^2} - \la O \ra + {\la O^{\prime}\ra\over M^2}, 
\eea
where $v_0=(1-\mn/q_0^2)^{1/2}$.  Note that the contribution of the
short cut is nonvanishing in this limit only in $T_l$.

As they stand, the above sum rules cannot be well saturated, since the
$2\pi$ and the continuum contributions on the spectral side are
missing. However, if we subtract the corresponding vacuum sum rules,
the continuum contribution will practically drop out.  We now go to
the limit of zero temperature, when the $2\pi$ contribution reduces to
zero. With $\mu > 0$, the nucleon and the antinucleon distribution
functions reduce in this limit to a $\th$-function and zero,
respectively. We thus finally arrive at the following sum rules at
finite nuclear density:
\bea
\overline{F^{\star 2}_\rho e^{-m_\rho^{\star 2}/M^2}} + 
{1\over {24 \pi^2}} \int_{\mn}^{4 \mu^2} ds (1-e^{-s/M^2})
\sqrt{1-\mn/s}(1+2m_N^2/s)
& =& {\OP\over M^2} - {\OPp \over {2M^4}} , \label{SR1} \\ 
\overline{m_{\rho}^{\star 2}F^{\star 2}_\rho e^{-m_\rho^{\star 2}/M^2}} - 
{1\over {24 \pi^2}} \int_{\mn}^{4 \mu^2} ds \, s \,  
e^{-s/M^2}\sqrt{1-\mn/s}(1+2m_N^2/s) 
& =& -\OP + {\OPp\over M^2} , \label{SR2} 
\eea
where the bar over a quantity denotes subtraction of its vacuum value,
e.g., $\overline{\la O \ra} = \la O \ra -\la 0|O|0\ra.$

The spectral integrals in the sum rules can be expanded in powers of
the Fermi momentum $p_F =\sqrt{\mu^2-m_N^2},$ or the nucleon number
density \cite{Jin}
\be
\bar{n} =4\int {d^3p\over {(2\pi)^3}} \th (p_F-|\vec{p}|) =
{2p_F^3\over {3\pi^2}} . 
\ee
The operator matrix elements can also be expanded in powers of the
nucleon number density. With the normalization of the one nucleon
state $\la p, \al | p^{\prime}, \al^{\prime}\ra =(2\pi)^3 2p_0
\de_{\al ,\al^{\prime}} \de^3 (\vec{p}-\vec{p}^{\prime}), \al,
\al^{\prime} $ denoting the spin and isospin degrees of freedom, we
have the expansion for any operator $R$ up to first order as
\be
\la R \ra = \la 0|R|0\ra + \int {d^3p\over {(2\pi)^3 2p_0}} \sum_\al 
\la p,\al |R|p,\al \ra \th (p_F-|\vec{p}|) . 
\ee
Denoting the constant, averaged matrix element by $\la p|R| p\ra
$, we get 
\be
\la R\ra = \la 0|R| 0 \ra +{\la p|R|p\ra\over {2m_N}} \bar{n} . 
\ee
Thus
\be
\la \hm\qq \ra = \la 0|\hm\qq |0\ra +\sg \bar{n},
\ee
where the $\sg$ matrix element is defined by 
$\sg = \la p|\hm(\bar{u}u +\bar{d} d)| p\ra/(2m_N) \simeq 45~{\rm MeV}$
\cite{Gasser}. The two-gluon operator is determined by the trace
anomaly
\be
\Th_{\mu}^{\mu} = -{9\over 8} {\al_s\over \pi} G_{\mu\nu}G^{\mu\nu} +
\hm\qq .
\ee
We then get 
\be
\la G^2\ra = \la 0 | G^2| 0\ra -{8\over 9} (m_N-\sg) \bar{n} . 
\ee
To determine the quark and the gluon energy densities in the medium, we need
their nucleon matrix elements. More generally one may write the
traceless $\Th_{\mu\nu}^{q,g}$ as 
\be
\la p|\Th_{\mu\nu}^{q,g} |p\ra =2  A^{q,g} \left(p_{\mu} p_{\nu}
-{1\over 4} g_{\mu\nu} p^2\right) .
\ee
The factor $2$ ensures that the constants satisfy $\sum_f A^f + A^g
=1$.  These constants individually may be obtained from fits with the
data on deep inelastic scattering. Based on the parametrizations for
the quark and the gluon distribution functions in Ref.~\cite{MRST},
these are found at $Q^2 = 1~{\rm GeV}^2$ to be 
\be
A^q = 0.62, \qquad A^g = 0.35 .
\ee
Collecting the above results we get
\bea
\OP &=& C\bar{n}, \quad
C={\sg\over 2}-{1\over 27}(m_N -\sg) +{3\over 22}m_N
\left\{A^q + A^g +  \lm(M^2) \left({8 \over 3} A^q - A^g \right)
\right\}, \\ 
\OPp &=& D\bar{n}, \quad 
D={56\over 81} \pi\al_s \la 0| \qq |0\ra {\sigma\over \hm},
\eea
where the quark condensate in vacuum is given by $\la 0|\bar{q}
q |0\ra / 2 = \la 0|\bar{u} u |0\ra =-(225~{\rm MeV})^3$. Furthermore, we
will use $\hm = 7~{\rm MeV}$ below. 

The sum rules now determine the nuclear density dependence of the
parameters of the $\rho$ meson. To first order in the number density,
we may expand these as
\be \label{def_ab} 
m^{\star}_{\rho} = m_{\rho} \left( 1+a{\bar{n}\over \bar{n}_s} \right),
\qquad 
F^{\star}_{\rho} = F_{\rho} \left( 1+b{\bar{n}\over \bar{n}_s} \right),
\ee 
where $\bar{n}_s$ is the saturation nucleon number density, $\bar{n}_s
= (110 \ {\rm MeV})^3$. From the sum rules (\ref{SR1}) and (\ref{SR2})
we then obtain the following expressions for the coefficients $a $ and
$b$:
\bea
a &= &-{\bar{n}_s \over 2 \frs} e^{\mrs/M^2} \Bigg[ C \left( {1\over
\mrs}+{1\over M^2} \right) 
-{D\over M^2} \left( {1\over \mrs} + {1\over {2M^2}}\right) 
-{1\over{4m_N}} -\left({m_N\over \mrs} -  {1\over {4m_N}} \right)
e^{-\mn/M^2} \Bigg] , \\
b &= & - {\bar{n}_s \over 2 \frs} e^{\mrs/M^2} \Bigg[ C {\mrs\over M^4} 
-{D\over {2M^4}} \left( 1+{\mrs\over M^2} \right) 
+ {1\over {4m_N}}\left( 1-{\mrs\over M^2} \right) - \left({m_N\over
{M^2}} + {1\over {4m_N}}\left( 1-{\mrs\over M^2} \right) \right)
e^{-\mn/M^2}\Bigg] . 
\eea

\begin{figure}[!h]
\centerline{\psfig{figure=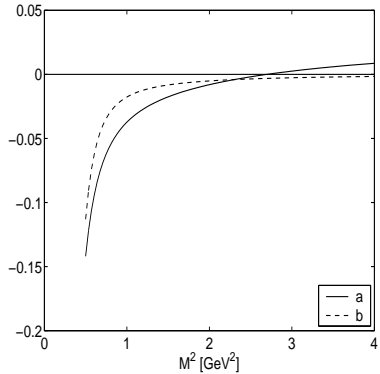,height=5cm,width=5cm}}
\caption{Coefficients of $\bar{n} / {\bar{n}_s}$ in the expansion of 
$m_{\rho}^{\star}$ and $F_{\rho}^{\star}$ from Eq.~(\ref{def_ab}) as
function of the Borel parameter $M$.} \label{fig:plot_a_b}
\end{figure}

A numerical evaluation of $a$ and $b$ is shown in
Fig.~\ref{fig:plot_a_b}. It will be noted that there is no sign of
constancy of $a$ and $b$ in any region of $M^2$; rather, they increase
rapidly and monotonically throughout, in particular $a$.  We conclude
that the sum rules as saturated above are inconsistent and cannot give
any reliable information about the density dependence of $m_{\rho}$
and $ F_{\rho}$.


\section{Discussion}
\label{sec:discussion}

We have studied QCD sum rules in the nuclear medium in the
$\rho$-meson channel using a saturation scheme considered earlier by a
number of authors \cite{Hatsuda,Jin_Leinweber,Cohen}. The sum rules we
actually evaluate consist of terms, all proportional to the nuclear
density. We find that the results are not stable in any range of
variation of the Borel parameter.  We conclude that the sum rules with
the present saturation scheme become inconsistent, in apparent
contradiction with the evaluation by these authors.

The resolution of this contradiction lies in the fact that the ``full''
sum rules considered by these authors are dominated by vacuum
contributions, the density dependent terms serving as small
perturbations.  As stated already in the Introduction, their stability
follows essentially from that of the vacuum sum rules. The variation
of some free parameters such as $s_{0}$, the beginning of the continuum,
in these works would further contribute to this apparent stability.

It would appear from earlier works that QCD sum rules can accommodate
different saturation schemes, each one giving rise to a new set of
results \cite{Leupold1}.  By working with a more sensitive version of
the sum rules, we show that the situation may not actually be so: a
simple saturation scheme, such as here, is rejected and a more
realistic one is called for.

Let us finally discuss the neglected contributions, whose inclusion
should restore the stability of these sum rules. At dimension 6,
five more operators would have contributed. These contributions can be
evaluated using the data on deep inelastic scattering \cite{Leupold2}.
To get an idea of the convergence of the series of operators, we note
that the contributions of a family of similar operators, differing
only in the number of Lorentz indices, form a series in powers of
$(m_N/M)$ with decreasing coefficients, given by the corresponding
moments of the structure functions. 
  
In the spectral function, it is not just the $N\bar{N}$ state, which
can give rise to the Landau cut in the low $q^2$ region.  Any
other state $N \bar{N}^{\star}$, where $N^{\star}$ is a resonance
communicating with the $\rho N$ channel, can contribute a cut for
$q^2\leq (m_{N^{\star}} -m_N)^2$. The two resonances $\Delta$(1232)
and N(1440) would presumably contribute significantly. To include
these contributions we need, however, know the
current-$N\bar{N}^{\star}$ couplings.  Also the increased width of the
$\rho$ meson in the medium is another source of altered contributions
to the sum rules.  We note that a study of such ``subtracted'' sum rules
for $\rho$ mesons at finite temperature showed no such instabilities
as mentioned above~\cite{Mallik_Mukherjee_2}.


\section*{Acknowledgments}

We thank H.\ Leutwyler and P.\ Minkowski for helpful discussions.  We
also acknowledge the kind hospitality at the Institute for Theoretical
Physics, University of Berne, where we started this work. This work
was supported in part by Schweizerischer Nationalfonds.


\end{document}